# Demonstration of liquid crystal for barocaloric cooling application

Zhongjian Xie[1], Yao Zhu

## Abstract
Current vapor-compression technology is based on the gas-liquid transition of hazardous gas. The alternative cooling technology focuses on the solid caloric material. A new liquid barocaloric material, i.e. the liquid crystal, is proved to be potential for cooling application primarily (based on other literatures). Its phase transition needs a small stress (tens of MPa) and can be operated near room temperature. It has a large entropy change of ~200 $J.K^{-1}.kg^{-1}$ (0.2 $J.K^{-1}.cm^{-1}$). There is no breakage problem due to its liquid form. It will provide the advantage for some unique cooling application cases comparing with solid caloric material. Moreover, this material is sustainable and non-toxic, which can meet the original intention of the alternative cooling technology.

## Introduction

Current vapor-compression technology is based on the gas-liquid transition of hazardous gas. The alternative cooling technology focuses on the solid caloric material. Then, what about the liquid caloric material?

Currently, all the investigations on mechanocaloric (mC) cooling devices are based on the elastocalorc (eC) effect of shape memory alloys (SMAs).[1–3] The devices based on the barocaloric (BC) effect have not yet been reported. Uniaxial stress is easier applied whereas hydrostatic pressure needs to be applied in all axes. For eC effect, the positions for the application of uniaxial stress and for the heat exchange can be separate, which facilitates the design of the device. For solid BC material, the application position of hydrostatic pressure conflicts with the position of the heat exchange. This may be difficult for designing a cooling device. The liquid BC material is thus proposed with three advantages compared to the solid one. Firstly, the problem of position conflict of heat transfer and pressure in solid BC material can be solved. The pressure can be just applied in one axis, and then all the liquid material will get a same hydrostatic pressure. The area for heat exchange can be in transverse axis. Secondly, there is no breakage problem for the liquid BC material, whereas it is a drawback for the solid mC material. Thirdly, the gas coolant in vapor compressor may be directly replaced by the liquid BC material with the least technology transformation.

In the investigation for cooling device based on eC effect of hard SMAs, the problem of large stress is prominent. Thus, the soft mC materials with a low stimulus stress/pressure need to be found. Moreover, the compressor in current refrigerator normally works at around 1 MPa,[4] which has already made a noise. Thus, larger stress/pressure for potential coolant (mC material) is not expected.

In this letter, in order to meet the needs of liquid BC material and small stimulus pressure for cooling device, the BC effect of liquid crystal is proposed and is proved based on other literatures.

## Direct measurement of barocaloric temperature change in liquid crystal

For cooling application, how can we learn from nature? There is no direct cooling example in nature. However, caloric effect may occur accompanying with structure transition. For living organism, the structure transition occurs in response to environmental change. Deep sea organisms, which are known as piezo- or baro-philic, have the mechanism to adapt their lipid membranes to maintain their fundamental fluid bilayer structure and mechanical properties in high pressure.[5,6] The main transition in lipid membrane is the gel/liquid crystalline transition. In the gel phase, the lipid molecule extends in the direction of membrane thickness and shrinks in the direction of membrane parallel, whereas the situation is reversed in the liquid crystalline phase.[7] Thus, the application of pressure can induce the gel/liquid crystalline transition.

---

[1] zhongjian.xie521@gmail.com

A lipid is a solid crystal with a well-ordered lamellar structures.[7] When adding water to the solid crystal, the liquid crystal is produced. One typical liquid crystal is dihexadecylphosphatidylethanolamine (DHPE) in excess water. Its BC effect was measured directly by Mencke *et al.*.[8] This measurement is for testing a pressure apparatus. The temperature variation and diffracted intensity by using time-resolved x-ray diffraction (TRXRD) are shown in Fig. 1. When the pressure is applied and removed, the temperature increases and decreases, respectively. It exhibits a conventional BC effect. It is attributed to the transition between the lamellar liquid crystal, $L_\alpha$, and the lamellar gel, $L_{\beta'}$, phase.

The pressure variation is 400 bar (40 MPa) and the temperature variation is ~3 K. The pressure is much lower than the current BC material (0.1-0.5 GPa) and the temperature variation is almost the same level (1-4.5 K).[9,10] The BC strength of this liquid crystal can be up to 75 K/GPa, which is much larger than current BC materials (4-15 K/GPa).[10]

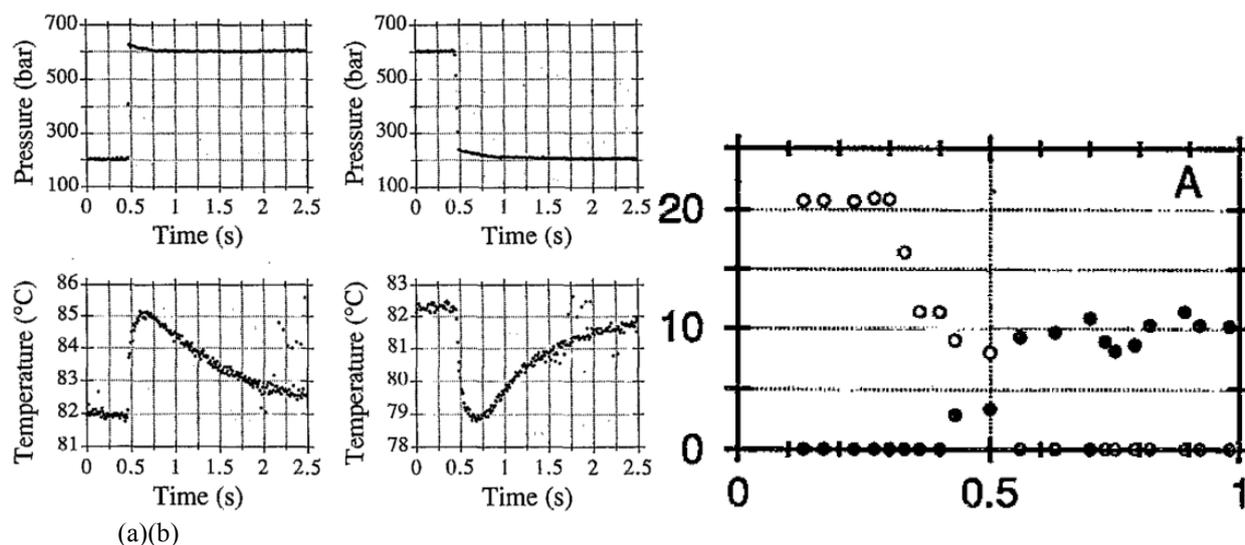

(a)(b)

fig. 1 (a) Typical pressure variation (200-600 bar) and the accompanying temperature variations for lipid, DHPE, in excess water. The temperature variation is from the enthalpy of the lamellar liquid crystal ($L_\alpha$)/gel ($L_{\beta'}$) transition. (b) The variation of diffracted intensity shows the $L_\alpha$ (hollow)/$L_{\beta'}$ (solid) phase transition acquired simultaneously with the pressure jump.[8]

The dynamic response of DHPE to pressure change was also tested by Mencke *et al.*.[8] The material is subjected to a sinusoidal pressure oscillation at 1 Hz. The dynamic TRXRD data on the phase transition is shown in Fig. 2. The diffracted intensity of two phases shows the fast response to pressure change. It provides insights into the kinetics and mechanism of the transition.[11] The liquid crystal can be used for a high-frequency cooling application. Moreover, it shows the reversibility of phase transition in a few pressure cycles.

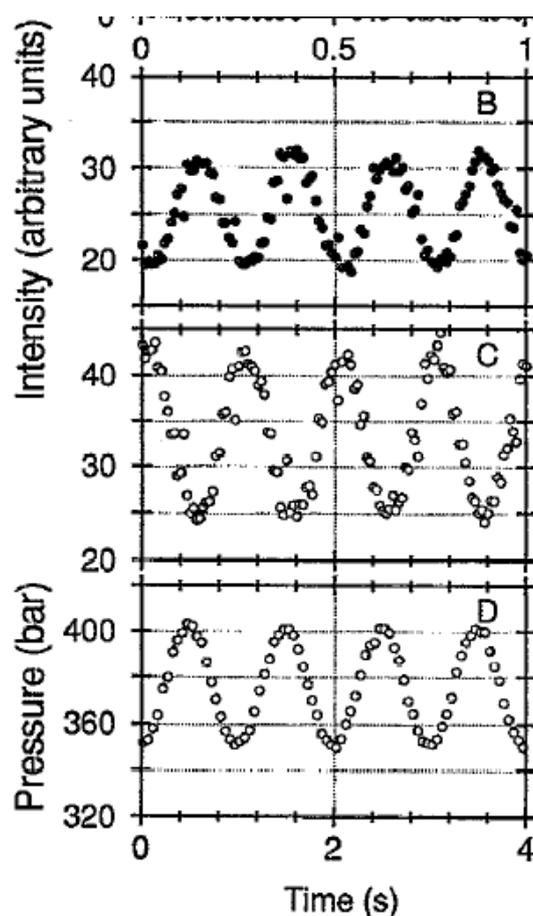

Fig. 2 TRXRD data acquired on a pressure-induced mesophase transition. The sinusoidal pressure oscillation is 1 Hz and peak-to-peak amplitude is 50 bar. It shows the relative amounts of lamellar liquid crystal (hollow)/gel (solid) transition in response to the oscillating pressure perturbation.[8]

For DHPE, one of the problems is the high transition temperature (82 °C), which limits its near room temperature cooling application. The transition temperature locating near room temperature can be realized in other lipid materials.

### Enthalpy, entropy change and transition temperature in liquid crystal

The bilayer phase-transitions of a series of 1,2-diacylphosphatidylcholines containing linear saturated acyl chain of carbons number (12, 13, 14, 15, 16, 17 and 18) was investigated by Ichimori et al.[12] When the carbon number of acyl chain increases from 12 to 18, the transition temperature increases from -2.1 °C to 55.6 °C and the entropy change increases from 28 to 137 $J.K^{-1}.mol^{-1}$ (Table 1). Considering the molecular weight of ~700 g/mol, the entropy change increases from 40 $J.K^{-1}.kg^{-1}$ to 200 $J.K^{-1}.kg^{-1}$. Thus, the acyl chain length can be used to adjust both the transition temperature and entropy change.

Table 1 Thermodynamic properties of phase transition for the bilayer membranes of diacylphosphatidylcholines[12]

| Lipid | Transition temperature | | $dT/dp$ | $\Delta H$ | | $\Delta S$ | $\Delta V$ |
|---|---|---|---|---|---|---|---|
| | (K) | (°C) | (K $MPa^{-1}$) | (kJ $mol^{-1}$) | (kcal $mol^{-1}$) | (J $K^{-1}$ $mol^{-1}$) | ($cm^3$ $mol^{-1}$) |
| 12:0-PC | 271.1 | −2.1 | 0.200 | 7.5[a] | 1.8[a] | 28 | 5.5 |
| 13:0-PC | 286.8 | 13.6 | 0.210 | 16.0 | 3.8 | 56 | 11.7 |
| 14:0-PC | 297.1 | 23.9 | 0.212 | 24.7 | 5.9 | 83 | 17.6 |
| 15:0-PC | 307.0 | 33.8 | 0.215 | 30.3 | 7.2 | 99 | 21.2 |
| 16:0-PC | 315.2 | 42.0 | 0.220 | 36.4 | 8.7 | 115 | 25.4 |
| 17:0-PC | 322.1 | 48.9 | 0.224 | 41.4 | 9.9 | 129 | 28.8 |
| 18:0-PC | 328.8 | 55.6 | 0.230 | 45.2 | 10.8 | 137 | 31.6 |

The transition temperature can also be shifted by pressure. In Fig. 3, as the pressure increases, the transition temperature increases. The pressure dependence of transition temperature, *dT/dP*, *T* is transition temperature and *P* is pressure, is ~ 0.2 K/MPa for different acyl chain-lengths (Table 1). This can be used to adjust the transition temperature at the hot end and cold end for a cooling system. At the cold end, a low transition temperature is needed and a low pressure can be applied. At the hot end, a high transition temperature is needed and a high pressure can be applied.

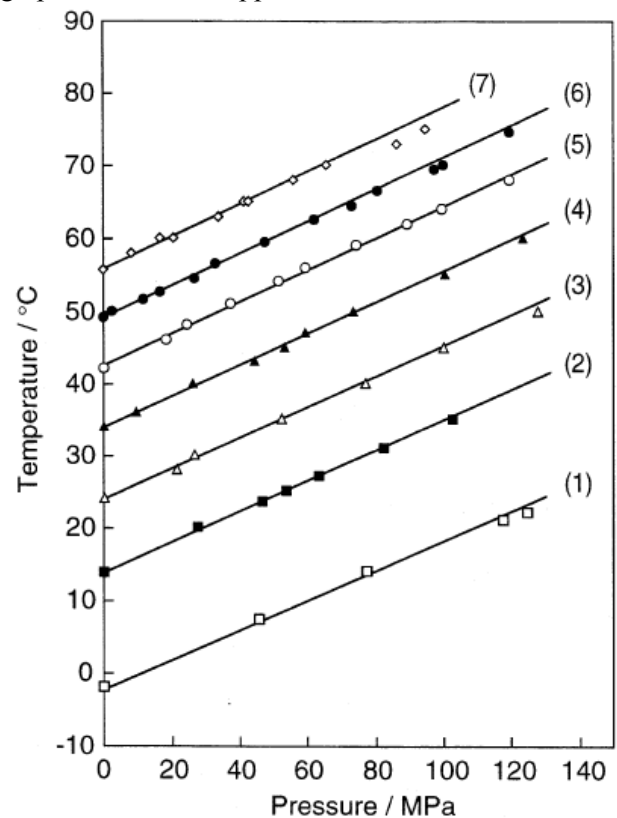

Fig. 3 Temperature—pressure phase boundaries between the ripple gel and liquid crystal phases for bilayer membranes of lipids with different acyl chain-lengths. (1) 12:0-PC, (2) 13:0-PC, (3) 14:0-PC, (4) 15:0-PC, (5) 16:0-PC, (6) 17:0-PC and (7) 18:0-PC.[12]

In Table 2, the liquid crystal is compared with other mechanocaloric materials, including Hydrochlorofluorocarbons (HCFC), shape memory alloys (SMAs), natural rubber (NR) and PVDF-based polymers. The stress of liquid crystal is the same level with PVDF-based polymers,[13] larger than HCFC[14] and natural rubber[15] while much smaller than Ni-Ti[16,17] and Cu-Zn-Al[18] alloys. The volumetric entropy change of liquid crystal is smaller than Ni-Ti but larger than other materials. The mass entropy change and enthalpy of liquid crystal are smaller than HCFC but larger than other solid materials.

Table 2 Comparison of liquid crystal with other mechanocaloric materials (all the values are estimated)

|  | HCFC[14] | Ni-Ti[16,17] | Cu-Zn-Al[18] | Natural rubber[15] | PVDF-based polymers[13] | Liquid crystal[12] |
|---|---|---|---|---|---|---|
| **Critical stress (MPa)** | 1 | 500 | >100 | 1 | tens | tens |
| **Volumetric entropy change (J.K$^{-1}$.cm$^{-3}$)** | 0.03 | 0.26 | 0.16 | 0.07 | 0.14 | 0.2 |
| **Mass entropy change (J.K$^{-1}$.kg$^{-1}$)** | 500 | 40 | 21 | 80 | 80 | 200 |
| **Enthalpy (kJ.kg$^{-1}$)** | 200 | 12 | 6 | 24 | 24 | 64 |

## Conclusion

In conclusion, the barocaloric effect of liquid crystal is proposed for cooling application. It needs a low pressure (tens of MPa). Its phase transition contributes to a large entropy change of ~200 J.K$^{-1}$.kg$^{-1}$, which is larger than solid mechanocaloric (mC) materials, including shape memory alloys (SMAs), natural rubber (NR) and PVDF-based polymers. Moreover, its phase transition can be performed near room temperature and can be adjusted by the acyl chain length and pressure.

Comparing with other solid mC materials, the absolute advantage of liquid crystal is its liquid form and it can be used for some unique cooling application case. Moreover, the liquid crystal has no breakage problem like solid mC materials. Most importantly, the material is sustainable and non-toxic.

Further detailed research of liquid crystal towards cooling application is needed. Moreover, some liquid crystals (ferroelectric liquid crystals) has been observed to possess the electrocaloric effect. The multicoupling method may be used to improve the caloric performance (multicaloric effect) of liquid crystal.

A bold idea is the biological refrigeration. The deep-sea bacteria with lipid membranes may be employed to perform the caloric effect and refrigeration. The advantage of biological refrigeration is its self-healing mechanism.